\documentclass{article}

\usepackage[final, nonatbib]{nips_2018}

\usepackage[utf8]{inputenc} 
\usepackage[T1]{fontenc}    
\usepackage[colorlinks=true, allcolors=blue]{hyperref} 
\usepackage{url}            
\usepackage{booktabs}       
\usepackage{amsfonts}       
\usepackage{nicefrac}       
\usepackage{microtype}      
\usepackage{color}
\usepackage{subfigure}
\usepackage{wrapfig}

\usepackage{amssymb}
\usepackage{mathtools}

\DeclarePairedDelimiter{\abs}{\lvert}{\rvert} 
\DeclarePairedDelimiter{\norm}{\lVert}{\rVert}
\usepackage{afterpage}

\newtheorem{example}{Example}

\title{Steerable Wavelet Scattering for 3D Atomic Systems with Application to Li-Si Energy Prediction}

\author{
  Xavier Brumwell\thanks{Contributed equally} \\
  Department of CMSE\thanks{Computational Mathematics, Science \& Engineering} \\
  Michigan State University\\
  \texttt{brumwel1@msu.edu} \\
  \And
  Paul $\text{Sinz}^{\ast}$ \\
  Department of $\text{CMSE}^{\dagger}$ \\
  Michigan State University\\
  \texttt{sinzpaul@msu.edu} \\
  \And
  Kwang Jin Kim\\
  Department of ChEMS\thanks{Chemical Engineering and Materials Science} \\
  Michigan State University \\
  \texttt{kimkwa12@msu.edu} \\
  \And
  Yue Qi\thanks{\href{https://researchgroups.msu.edu/msce}{\texttt{https://researchgroups.msu.edu/msce}}} \\
  Department of $\text{ChEMS}^{\ddagger}$ \\
  Michigan State University\\
  \texttt{yueqi@egr.msu.edu} \\
  \And
  Matthew Hirn\thanks{Corresponding author; \href{https://matthewhirn.com/}{\texttt{https://matthewhirn.com/}}} \\
  Department of $\text{CMSE}^{\dagger}$ \\
  Department of Mathematics\\
  East Lansing, MI 48864\\
  \texttt{mhirn@msu.edu} \\
}

\begin{document}

\maketitle

\begin{abstract}
  A general machine learning architecture is introduced that uses wavelet scattering coefficients of an inputted three dimensional signal as features. Solid harmonic wavelet scattering transforms of three dimensional signals were previously introduced in a machine learning framework for the regression of properties of small organic molecules. Here this approach is extended for general steerable wavelets which are equivariant to translations and rotations, resulting in a sparse model of the target function. The scattering coefficients inherit from the wavelets invariance to translations and rotations. As an illustration of this approach a linear regression model is learned for the formation energy of amorphous lithium-silicon material states trained over a database generated using plane-wave Density Functional Theory methods. State-of-the-art results are produced as compared to other machine learning approaches over similarly generated databases.  
\end{abstract}

\section{Introduction}

Modeling the behaviors of complex multiscale materials has become a domain of ever increasing research interest bringing together experts from a variety of fields ranging from chemistry, biology, and physics, to experts in numerical methods, both experimental and theoretical. The multiscale nature of such materials implies a dependence of material properties upon multiple length and/or time scales. This leads to several fundamental challenges such as the curse of dimensionality, in which the discretization of a material grows exponentially with system size. The coupling of scales is frequently nonlinear, meaning that it is not sufficient to simply separate length scales in computations and recombine the results in a linear fashion. The challenge of multiscale materials modeling is therefore to create models of phenomena of interest which include the influence and coupling of multiple scales in length and time while maintaining acceptably low computational costs. In the computation of energies of atomic systems, for example, plane-wave density functional theory (DFT) provides state-of-the art prediction capabilities for computing the ground state energy of a periodic atomic system; however, the computational costs of DFT restrict its practical usage to periodic systems on the order of a few hundred atoms \cite{B517914A}. This has led to the development of novel numerical methods with the chief objectives being computational speedups and accuracies that are competitive with the state-of-the-art methods (e.g., DFT). Machine learning models trained on datasets constructed with trusted methods have seen growing interest in recent years \cite{
chemicalDesign2018,
De:moleculesAcrossAlchemicalSpace2016,
schutt:molecuLeNet2017,
gilmer:MPNNQC2017,
montavon:mlMolProp2013,
burke:bypassingKohnSham2017,
PhysRevLett.114.096405,
fletcher2014prediction,
rupp2012fast}.
We introduce one such model as the focus of this work. 

We present a general machine learning architecture for the modeling of multiscale properties of complex materials. Any such machine learning algorithm must address the fundamental challenges in materials modeling, including the multiscale nature of many materials in addition to properties of invariance and stability. The machine learning architecture proposed here consists of a linear regression over a new type of three dimensional wavelet scattering coefficients \cite{mallat:scattering2012}, which encode localized fluctuations of signals that represent the atomistic state of the system. The scattering transform has an architecture similar to that of convolutional neural networks (CNNs), but replaces learned, fixed scale filters with a set of designed, multiscale wavelets. The resulting transform thus not only separates the scales of the atomistic system, but also extracts complex coefficients that incorporate geometric patterns at multiple scales. In order to obtain a rotationally invariant representation of the state of the atomistic system, this new scattering transform is constructed out of a general class of three dimensional steerable wavelets, which can be considered as generalizations of two dimensional steerable filters \cite{freeman:steerableFilters1991}. Within this framework, we additionally introduce a specific family of wavelets, which we refer to as atomic orbital wavelets, that generalizes the family of solid harmonic wavelets introduced in \cite{eickenberg:3DSolidHarmonicScat2017}.

Using the wavelet scattering coefficients, we train a linear regression model for predicting the formation energy of amorphous lithium-silicon (Li-Si) systems. The resulting model has an MAE of $0.78\pm 0.01 \text{meV/atom}$ and an RMSE of $1.24\pm 0.05 \text{meV/atom}$. Formation energies are the fundamental driving force for Li storage in Li-Si systems, which is of special interest for next generation high energy Li-ion batteries. Lithiation and delithiation of Si have been widely studied using reactive force field based MD simulations~\cite{Qi:vacanciesInSi2015,Qi:atomisticSimuation2017}. However, constructing interatomic potentials (or force fields) is a repetitive task of training on extensive first-principles calculations, generally resulting in several months of laborious iterative fitting. Furthermore, computed force fields do not generalize well due to the complex interdependence of the parameters. A paradigm shift is taking place in which machine learning potentials are rising as an alternative to harness first-principles accuracy and force field efficiency \cite{Artrith:LiSiML2018,
onat:implantNNLiSi2018,Seko2018,chmiela:MLconservativeFF2017,Behler:atomisticSimulations2016,smith:NNpotential2017,hansen:BoB2015, bartok:gaussAppPot2010,Behler:NNPotEnergy2007}. 

The remainder of this paper is organized as follows. In section \ref{sec:wavelets} we introduce a general class of three dimensional, steerable wavelets and the resulting equivariant scattering transform. Invariant scattering coefficients are derived in section \ref{sec:scattering}. The Li-Si database is described in section \ref{sec:materials}, and in section \ref{sec:numerics} numerical regression results on the formation energies of these systems, in addition to further analysis, are presented. Section \ref{sec: conclusion} contains some concluding remarks.

\section{Equivarient scattering networks with steerable wavelet filters} \label{sec:wavelets}

\subsection{Periodic steerable wavelets}

We introduce a general class of 3-dimensional steerable wavelet filters for which the resulting wavelet transform is equivariant with respect to translations and rotations when combined with an appropriately chosen nonlinearity. The equivariance of the transform leads to the associated invariance of the scattering coefficients introduced in section~\ref{sec:scattering}. These wavelet filters have the form
\begin{equation}\label{eq:generalwavelet}
    \Psi_{\gamma l}^m(u) = Q_{\gamma}(r) Y_l^m\left(\phi, \theta \right), \quad (r, \phi, \theta) \in \mathbb{R}^3,
\end{equation}
for an arbitrary radial function $Q_{\gamma}$ and the well-known spherical harmonic functions $Y_l^m$, indexed by parameters $l=0,1,2,\ldots$ and $-l\leq m \leq l$. Here we use spherical coordinates with $r$ the radial component, $\phi\in[0,2\pi]$ the azimuthal angle, and $\theta\in[0,\pi]$ the polar angle. While equivariance can be obtained for any $Q_{\gamma}$ (see section \ref{sec: equivariance}), we constrain $Q_{\gamma}$ so that the resulting function $\Psi_{\gamma l}^m$ is either a wavelet or low pass filter. There is therefore great freedom of choice in the radial component. Guidance in choosing $Q_{\gamma}$ may come from knowledge of the application or the types of signals to be analyzed; it can also be learned from data in a task driven fashion. In example~\ref{ex:atomic orbital wavelets} below, as well as the numerical experiments reported in section \ref{sec:numerics}, we specify a class of radial functions that are useful for modeling atomic systems. In this case it is also beneficial (for computational efficiency) to choose $Q_{\gamma}$ so that the Fourier transform of the wavelet, $\widehat{\Psi}_{\gamma l}^m$, can be computed analytically.

Wavelets $\Psi_{\gamma l}^m \in L^2(\mathbb{R}^3)$ are filters with zero average, $\int_{\mathbb{R}^3} \Psi_{\gamma l}^m = 0$, and which are localized (they have fast decay) in space and frequency. A multiscale wavelet transform of a signal $\rho\in L^2(\mathbb{R}^3)$ is given by the convolution of the signal with dilations of the wavelets $\Psi_{\gamma lj}^m(u) = 2^{-3j} \Psi_{\gamma l}^m (2^{-j} u)$. For periodic signals $\rho \in L^2 (\mathbb{T}^3)$ (where $\mathbb{T}$ is the torus), which are used to model materials, we periodize the wavelets and compute circular convolutions. Periodized wavelets with period $P$ are defined as:
\begin{equation*}
    \psi_{\gamma lj}^m (u) = \sum_{k \in \mathbb{Z}} \Psi_{\gamma lj}^m (u - kP)
\end{equation*}
and circular convolutions over $\mathbb{T}$ are denoted by $\circledast$. A periodized wavelet can be obtained efficiently by appropriately sub-sampling its Fourier transform. The resulting wavelet coefficients are:
\begin{equation*} \label{eq:wavelet transform}
    W\rho = \{\rho \circledast \psi_{\gamma lj}^m(u) : \gamma \in \Gamma, \,  j\in \mathbb{Z}, \, -l \leq m \leq l, \, 0\leq l \leq L\}.
\end{equation*} 

\begin{wrapfigure}[22]{R}{0.41\linewidth}
   \centering
    \includegraphics[width=0.3\linewidth]{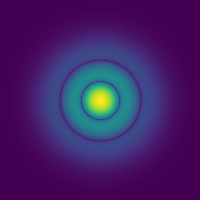}
    \includegraphics[width=0.3\linewidth]{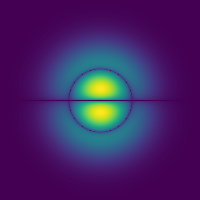}
    \includegraphics[width=0.3\linewidth]{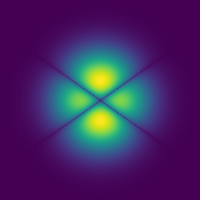}\\
    \includegraphics[width=0.3\linewidth]{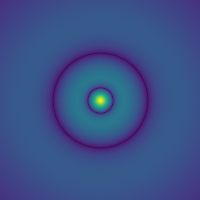}
    \includegraphics[width=0.3\linewidth]{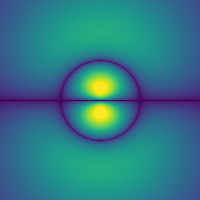}
    \includegraphics[width=0.3\linewidth]{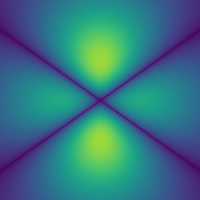}
   \caption{Top row left to right: Density plot cross sections in the $xz$-plane of atomic orbital wavelets for $(n,l,m)=(3,0,0)$, $(3,1,0)$, and $(3,2,0)$.  Bottom Row: Corresponding plots for the hydrogen atom orbitals  ($3s$, $3p$, and $3d$ orbitals, respectively). All images are rescaled for visualization. Note that the exponential radial decay $e^{-r}$ of the hydrogen orbitals is replaced by a Gaussian decay $e^{-r^2}$ in the wavelets accounting for the greater localization of the wavelets.}
   \label{fig:densityPlots}
\end{wrapfigure}
Choosing wavelets with simple Fourier transforms allows for $W\rho$ to be computed in frequency and brought back to space via the Fast Fourier Transform rather than computing the convolution directly, allowing for considerable time savings. 

The wavelet transform can be viewed as an alternative to the Fourier transform. Whereas the Fourier transform extracts responses against complex sinusoidal waveforms with a continuum of frequencies along the entire domain of a signal, the wavelet transform uses wavelets to filter out behavior localized in space which is captured by the various dilations of the mother wavelets. 

For the present study we illustrate the flexibility of \eqref{eq:generalwavelet} by introducing wavelets where the radial component depends on $\gamma = (n,l)$, where $n$ is an extra parameter. In this way we may create multiple mother wavelets that independently vary in radial and angular directions as specified by the parameters $n$, $l$, and $m$. Example~\ref{ex:atomic orbital wavelets} below introduces a specific choice of $Q_{nl}$ that is designed for the regression of energies of atomic systems. We refer to these wavelets as atomic orbital wavelets. These wavelets are used in all scattering regressions, which are described in section \ref{sec:numerics}.

\begin{example}\label{ex:atomic orbital wavelets}
The atomic orbital wavelets have the form of  \eqref{eq:generalwavelet} with radial component
\begin{equation*}
    \,\,\,
    Q_{nl}(r) = K_{nl}(\alpha) r^l  L_{n-l-1}^{l+1/2}(r^2/(2\alpha^2))  e^{-r^2/(2\alpha^2)},
    \quad K_{nl\alpha} = \frac{2^{n-l-1}(n-l-1)!}{(\sqrt{2\pi})^{3}\alpha^{2n+1}}\sqrt{\frac{4\pi}{2l+1}}
\end{equation*}
for normalizing factor $K_{nl\alpha}$, the usual quantum numbers $n$, $l$, $m$, and parameter $\alpha$ corresponding to the standard deviation of the Gaussian which determines the locality of the mother wavelet. The Fourier transform has a simple form and is given by
\begin{equation*}\label{eq:atomic orbital wavelets fourier}
    \widehat\Psi_{nl}^{m}(\eta,\xi,\zeta) = (-i)^l \eta^{2(n-1)-l}e^{-\alpha^2\eta^2/2}\sqrt{\frac{4\pi}{2l+1}}Y_l^m(\xi,\zeta).
\end{equation*}
Here $L_k^\nu$ are the associated Laguerre polynomials defined by $L_{k}^{\nu}(x) = \left(x^{\nu} {k!}\right)^{-1}{(\frac{d}{dx}-1)^k}x^{k+\nu}$. The frequency coordinates are written in spherical coordinates with $\eta$ the radius, $\xi \in [0,2\pi]$ the azimuthal angle, and $\zeta\in[0,\pi]$ the polar angle.
These wavelets include the solid harmonic wavelets which are recovered for $l=n-1$ since $L_0^\nu = 1$ for any $\nu$. The solid harmonic wavelets are used for molecular energy regression in \cite{eickenberg:3DSolidHarmonicScat2017, eickenberg:scatMoleculesJCP2018}. 
\end{example}

The choice of $Q_{nl}$ in example~\ref{ex:atomic orbital wavelets} is made to mimic the geometry of the atomic orbitals of the hydrogen atom; these orbitals are closed form solutions to the time-independent Schr\"{o}dinger equation. We note that the solid harmonic wavelets of \cite{eickenberg:3DSolidHarmonicScat2017, eickenberg:scatMoleculesJCP2018} mimic the $1s, 2p, 3d, 4f, \ldots,$ orbitals of the hydrogen atom. This leaves out the majority of the orbitals which, however, the atomic orbital wavelets recover by mimicking the full set of orbitals: $1s, 2s, 2p, 3s, 3p, 3d, \ldots.$ 
The atomic orbital wavelets are a modification of the spherical Gaussian type orbitals used in \cite{kuang:SGTOs1997}, and references therein, where they are used for producing simple formulas for multi-centered molecular integrals. 

Cross sections of three atomic orbital wavelets are illustrated in figure~\ref{fig:densityPlots} along with plots of the corresponding orbitals associated to the hydrogen atom. Unlike atomic orbitals which are at a fixed scale, we utilize several dilations of the atomic orbital wavelets for use in the multiscale wavelet transform. In the next section we see that a nonlinear operator, similar to the modulus, applied to the wavelet coefficients is used to recover equivariance to rotations and reflections.

\subsection{Nonlinear equivariance} \label{sec: equivariance}

Translation equivariance is automatically guaranteed for any convolution operator, that is, for $t\in\mathbb{R}^3$, $f \circledast g(u-t) = f_t \circledast g(u)=f \circledast g_t(u)$ where $f_t(u)=f(u-t)$. Rotations of the spherical harmonics are distributed across all values of the parameter $m$ for a fixed $l$ as
\begin{equation*}
    Y_l^m(Ru) = \sum_{m'=-l}^{l} [D_{mm'}^{(l)}(R)]^* Y_l^{m'}(u)
\end{equation*}
where $[D_{mm'}^{(l)}(R)]^*$ is the complex conjugate of the $(m, m')$ entry of the Wigner D-matrix $D^{(l)}(R)$ associated to the rotation $R\in SO(3)$. It follows that the wavelet filters $\Psi_{\gamma l}^m$ of \eqref{eq:generalwavelet} are steerable. 

To recover rotationally equivariant coefficients we apply a nonlinear operator $\sigma$ (which we shall refer to as the modulus), first introduced in \cite{eickenberg:3DSolidHarmonicScat2017}, which is defined by
\begin{equation}\label{eq:wavelet modulus}
    \sigma(\rho\circledast\psi_{\gamma lj})(u)= \left(\sum_{m=-l}^l \abs*{\rho \circledast  \psi_{\gamma lj}^m(u)}^2\right)^{1/2}
\end{equation}
with a slight abuse of notation by omitting the summing variable $m$ on the left hand side. The resulting set of coefficients 
\begin{equation*}
    \sigma (W \rho)  = \{\sigma\left(\rho\circledast\psi_{\gamma lj}^m\right) : \gamma \in \Gamma, \, j\in \mathbb{Z},\, 0 \leq l \leq L\}
\end{equation*}
are equivariant to rotations, as follows from Uns\"old's Theorem and by noting the Wigner D-matrix is unitary; thus
$
    \sigma(\rho\circledast\psi_{\gamma lj})(Ru)=\sigma(\rho_R \circledast\psi_{\gamma lj})(u)
$
for any rotation $R \in SO(3)$ where $\rho_R(u) = \rho(Ru)$ is a rotation of the input signal. 
The rotational equivariance and steerability of this class of wavelets is entirely dependent upon the spherical harmonics; this modulus operator guarantees rotational equivariance regardless of the choice of $Q_{\gamma}$ in \eqref{eq:generalwavelet}. Furthermore, the modulus operator guarantees invariance to reflections as $Q_{\gamma}(-r)=Q_{\gamma}(r)$ and $Y_l^m(-\textbf{r})=(-1)^l Y_l^m(\textbf{r})$.

\subsection{Scale coupling through multiple layers}

The strength of \eqref{eq:generalwavelet} and \eqref{eq:wavelet modulus} lies not only in providing equivariance to rigid motions, but also in that the radial component can be tailor-made for a specific application. For multiscale problems choosing an appropriate radial component will allow for the separation of length scales of the input signal $\rho$. For example, the electronic density $\rho$ of an atomic system depends upon interactions of electrons with other nearby electrons and nuclei as well as interactions with groupings of atoms at longer length scales. Separating scales is advantageous for computational purposes; however, to adequately model a multiscale system, the inherent coupling between scales must be recovered.

A second wavelet modulus may be computed to recover interactions between length scales. For example, if a wavelet coefficient shows a response to a certain geometric feature of the input at a given scale $2^{j_1}$, then the cross-scale interactions of this feature will become visible to wavelet modulus coefficients at a larger scale $2^{j_2}$. Therefore we compute second order wavelet modulus coefficients 
\begin{equation}\label{eq:second wavelet modulus}
    \sigma(\sigma(\rho\circledast\psi_{\gamma_1 l_1j_1})\circledast\psi_{\gamma_2 l_2j_2})=\left(\sum_{m=-l_2}^{l_2} \abs*{\sigma (\rho\circledast \psi_{\gamma_1 l_1j_1})\circledast\psi_{\gamma_2 l_2j_2}^{m}(u)}^2\right)^{1/2}, \quad j_1<j_2,
\end{equation}
where we apply the modulus operator $\sigma$ at each level to maintain the equivariance established for the first order coefficients. First order wavelet modulus coefficients encode interference within a signal $\rho$ at the scale $2^{j_1}$. Applying a second order wavelet transform with the modulus operator then encodes the interference within the first order interference at the scale $2^{j_2}$, coupling these two scales. We may therefore compute a cascade of wavelet transforms, applying the nonlinear modulus operator at each level, and capture higher order interactions at each level
\begin{equation}\label{eq:multiorder wavelet modulus}
    \sigma(\sigma(\cdots\sigma( \rho\circledast\psi_{\gamma_1 l_1j_1})\circledast\psi_{\gamma_2 l_2j_2})\cdots\circledast\psi_{\gamma_k l_k j_k}), \quad {j_1 < \cdots < j_k}.
\end{equation}
The resulting set of coefficients at order $k$ is a subset of 
\begin{equation*}
    \underbrace{\sigma (W \sigma (W \cdots \sigma (W \rho)))}_{k \text{ times}} = \big\{ \sigma(\sigma(\cdots\sigma( \rho\circledast\psi_{\gamma_1 l_1j_1})\circledast\psi_{\gamma_2 l_2j_2})\cdots\circledast\psi_{\gamma_k l_k j_k}) \big\}_{\gamma_k, j_k, l_k}
\end{equation*}
which mimics the alternating cascade of linear and nonlinear transformations found in CNNs. Since each wavelet modulus operator $\sigma \circ W$ is equivariant to rotations, the entire cascade is equivariant. The transform \eqref{eq:multiorder wavelet modulus} is thus similar to an equivariant three dimensional CNN constructed out of steerable filters; related studies on equivariant neural networks have previously been undertaken for 2D CNNs \cite{pmlr-v48-cohenc16, cohen:steerableCNNs2016}, groups and graphs \cite{kondor:covariantCompNets2018, kondor:equivarianceNNGroups2018}, and 2D scattering transforms \cite{mallat:rotoScat2013}. 

\begin{figure}[hbtp]
    \centering
    \includegraphics[width=0.8in]{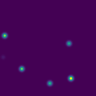}
    \includegraphics[width=0.8in]{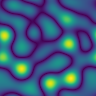}
    \includegraphics[width=0.8in]{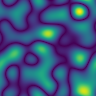}
    \includegraphics[width=0.8in]{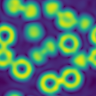}
    \includegraphics[width=0.8in]{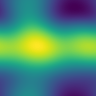}
    \caption{Cross-sectional views of a 3-dimensional signal $\rho$ and its wavelet modulus coefficients specified by parameters $(n,l,j)$ and $(n_1,l_1,j_1,n_2,l_2,j_2)$ for $\sigma(\rho\circledast\psi_{nlj})$ and $\sigma(\sigma(\rho\circledast\psi_{n_1 l_1 j_1})\circledast\psi_{n_2 l_2 j_2})$. From left to right they are $\rho$, $(2,0,2)$, $(2,0,2,5,0,2.5)$, $(5,2,1.5)$, and $(5, 2, 1.5, 5, 4, 2)$  Values range from dark purple at zero to yellow at large values.}
    \label{fig:waveletTransform}
\end{figure}

In this work we restrict computations to first and second order cascades of wavelet modulus transforms. Figure~\ref{fig:waveletTransform} shows a cross-section of a three dimensional signal $\rho$ that represents an atomic state, and the same cross-sectional view of the first and second order wavelet modulus coefficients. The wavelets used are those of example~\ref{ex:atomic orbital wavelets} and the input signal is a fictional electronic density given by a sum of normalized Gaussians scaled by electronic charges. The multiscale interactions are clearly visible. The second image from the left shows interactions between the five visible atoms in the cross-section along with interactions into and out of the page. The third image from the left contains second order coefficients of the prior image and shows similar but more complicated interference patterns. The fourth image shows perimeters of the atoms as well as connections between neighboring atoms and the final image is an associated second order coefficient which shows long length scale interference. We see here that different wavelets as specified by the parameters $(n,l,j)$ capture very different scale interactions, from fine scales which show the atomic locations clearly, to large length scales that show larger interactions.

\section{Invariant scattering coefficients}
\label{sec:scattering}

\subsection{Translation and rotation invariance}

We seek features that are \emph{invariant} to translations and rotations; however, the wavelet modulus coefficients \eqref{eq:wavelet modulus}, \eqref{eq:second wavelet modulus}, and \eqref{eq:multiorder wavelet modulus} are only \emph{equivariant} to these transformations. Invariant coefficients are recovered by summing the wavelet modulus coefficients giving the first and second order wavelet scattering coefficients as
\begin{gather}
    \norm*{\sigma(\rho\circledast\psi_{\gamma lj})}_q^q = \int_{\mathbb{R}^3} \abs*{\sigma(\rho\circledast\psi_{\gamma lj})(u)}^q \, du,\label{eq:first scattering coefficients} \\
    \norm*{\sigma\left(\sigma(\rho\circledast\psi_{\gamma_1 l_1 j_1})\circledast\psi_{\gamma_2 l_2 j_2}\right)}_q^q = \int_{\mathbb{R}^3} \abs*{\sigma\left(\sigma(\rho\circledast\psi_{\gamma_1 l_1 j_1})\circledast\psi_{\gamma_2 l_2 j_2}\right)(u)}^q \, du.\label{eq:second scattering coefficients}
\end{gather}
Here the wavelet modulus coefficients for a fixed $(\gamma, l,j,q)$ or $(\gamma_1, l_1,j_1,\gamma_2, l_2,j_2,q)$ are aggregated into a single number and are chosen as the features for prediction of properties of input signals $\rho \in L^2(\mathbb{T}^3)$. These coefficients are invariant to translations and rotations of $\rho$ and encode information at a fixed scale of $2^{j_1}$ in the first order and cross-scale features between $2^{j_1}$ and $2^{j_2}$ in the second order coefficients. Multiple integration factors $q$ do not effect the invariance of the coefficients and are utilized in order to emphasize different responses of the input. 

The specific integration factors may be chosen to correspond to interpretable phenomena. In the numerical experiments described in section \ref{sec:numerics} we select $q = 1, \nicefrac{4}{3}, \nicefrac{5}{3}, 2$. The power $q=1$ scales with the number of particles in the system, whereas it is shown in \cite{hirn:waveletScatQuantum2016} that first order scattering coefficients for $q=2$ encode pairwise interactions, which capture the electrostatic Coulomb interactions between electrons. Furthermore, one can show that $\| \rho \|_{\nicefrac{5}{3}}^{\nicefrac{5}{3}}$ combined with the collection $\| \sigma (\sqrt{\rho} \circledast \psi_{n(n-1)j}) \|_2^2$ capture the kinetic energy of the system, as approximated by the Thomas-Fermi-Dirac-von Weizsäcker (TFDW) model. Finally, a multiple of the $L^{\nicefrac{4}{3}}$ integral of $\rho$ is a first order approximation of the exchange energy; see again the TFDW model \cite{cances:handbook2003}.

While this analysis maps the scattering coefficients nicely onto the TFDW model, it is only an approximation of the true ground state energy of an atomistic state. Furthermore, there is information loss in the summing of the wavelet modulus coefficients. These considerations provide further motivation for including higher order scattering coefficients \eqref{eq:second scattering coefficients} as features in our regression model.

\subsection{System size invariance}

We seek to build a machine learning model to predict the formation energy per $\text{Li}_\beta \text{Si}_{1-\beta}$ formula unit of an infinitely large periodic atomic state; see section~\ref{sec:materials} below. The formation energy is independent of system size (number of atoms), and therefore our features must recover this additional invariant. Normalizing the input signals to the wavelet transform gives signals $\nicefrac{\rho}{\norm*{\rho}_q}$ and $\nicefrac{\sigma(\rho\circledast\psi_{\gamma lj})}{\norm*{\sigma(\rho\circledast\psi_{\gamma lj})}_q}$, which results in the following features that are invariant to system size:
\begin{equation}\label{eq:normalized scattering coefficients}
    \frac{\norm*{\sigma(\rho\circledast\psi_{\gamma lj})}_q^q}{\norm*{\rho}_q^q},
    \qquad
    \frac{\norm*{\sigma\left(\sigma(\rho\circledast\psi_{\gamma_1 l_1 j_1})\circledast\psi_{\gamma_2 l_2 j_2}\right)}_q^q}{\norm*{\sigma(\rho\circledast\psi_{\gamma_1 l_1 j_1})}_q^q}.
\end{equation}
These features are normalized versions of the features \eqref{eq:first scattering coefficients} and \eqref{eq:second scattering coefficients}.

\begin{wrapfigure}[17]{R}{0.45\linewidth}
  \centering 
  \includegraphics[trim={0 0.1in 0 0.2in},clip,width=0.99\linewidth]{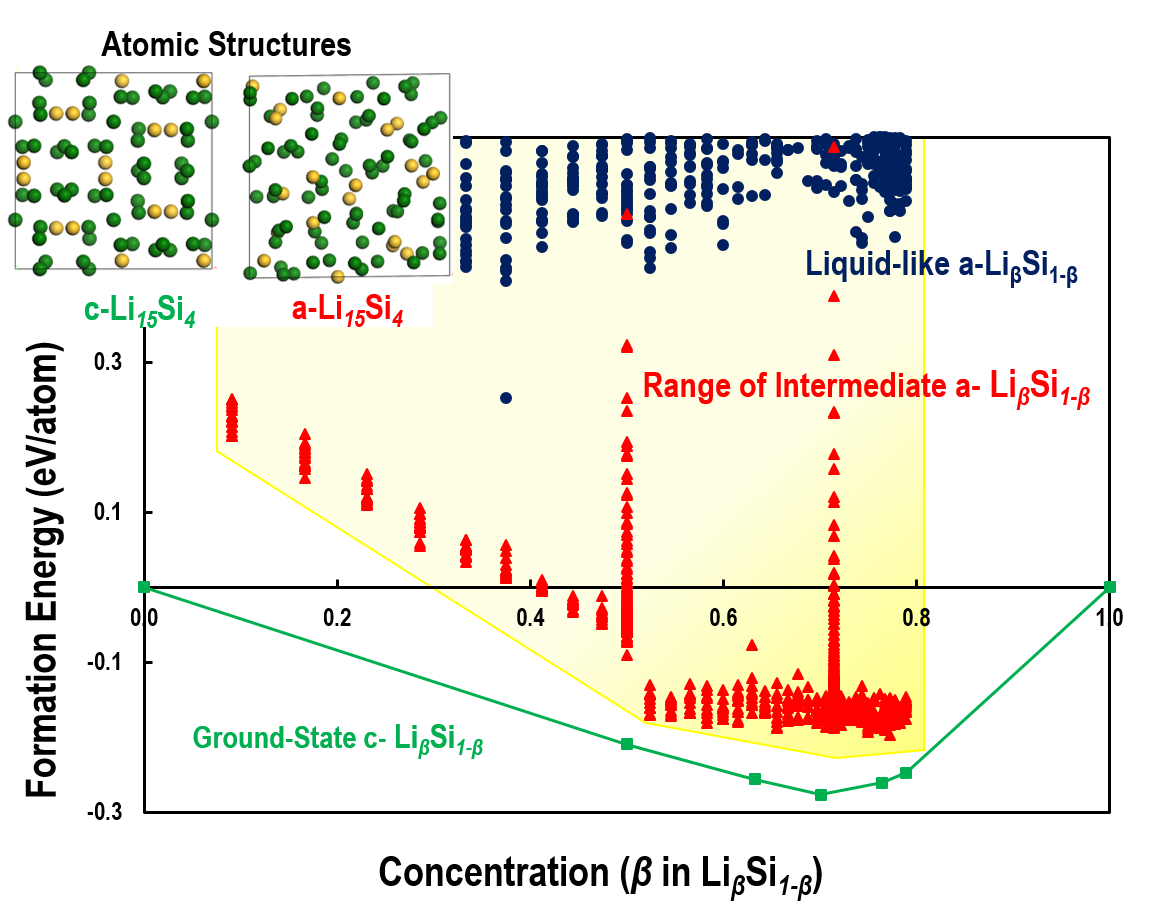}
  \caption{Formation energies of a-Li\textsubscript{$\beta$}Si\textsubscript{$1-\beta$}. Green crystalline states are not used but shown here for reference. In the atomic structures Li atoms are green and Si are yellow.}
  \label{fig:scatterPlot}
\end{wrapfigure}

Infinitely large amorphous systems are inherently disordered and, as such, can be modeled as a 3D compound Poisson process in space where the location of an atom is an arrival event of the process. In practice, plane-wave DFT simulations utilize a finite unit cell with periodic boundary that creates order within the system; however, this phenomena is mitigated by taking a large enough unit cell. In \cite{bruna:scatMoments2015}, general wavelets are used to compute scattering coefficients using a similar normalization to \eqref{eq:normalized scattering coefficients}. There it is shown that normalized first and second order scattering coefficients characterize regular Poisson processes in 1D. We can thus interpret these scattering coefficients \eqref{eq:normalized scattering coefficients} as statistical moments of $\rho$. We train a second order model using these normalized features on a database specified in section~\ref{sec:materials} and discuss the results of our numerical experiments in section~\ref{sec:numerics}.

\section{Database of Li-Si systems}
\label{sec:materials}

Let $x = \{ (z_k, r_k) \}_{k=1}^N$ denote the state of an atomistic system, in which $z_k$ denotes the total protonic charge of the $k^{\text{th}}$ atom and $r_k \in \mathbb{R}^3$ its position. In this paper we consider amorphous Li-Si systems \mbox{a-Li\textsubscript{$\beta(x)$}Si\textsubscript{$1-\beta(x)$}} for $0\leq \beta(x) \leq 1$ the atomic fraction of Li for state $x$. For notational clarity we write $\beta$ rather than $\beta(x)$ with the dependence on the state implied.
The formation energy per formula unit of an atomic system is defined relative to the total energy per formula unit of the system, $E (x)$, the per atom energy of crystalline Si, $E_{\text{Si}}$, and the per atom energy of body-centered cubic Li, $E_{\text{Li}}$. The formation energy of an a-Li\textsubscript{$\beta$}Si\textsubscript{$1-\beta$} state $x$ is computed as
\begin{equation}
    E_f (x)= E (x) -\beta E_{\text{Li}} - (1-\beta)E_{\text{Si}} .
\end{equation}

To generate an extensive training set that includes all local structural motifs and therefore covers the entire configuration space of a-Li\textsubscript{$\beta$}Si\textsubscript{$1-\beta$}, a wide range of a-Li\textsubscript{$\beta$}Si\textsubscript{$1-\beta$} structures with $\beta$ ranging from $0$ to approximately $0.79$ were prepared. Figure~\ref{fig:scatterPlot} contains a scatterplot showing the range of energies computed and two example Li-Si structures.
Liquid-like a-Li\textsubscript{$\beta$}Si\textsubscript{$1-\beta$} structures were generated for each of 37 concentrations ($\beta$) using ReaxFF-MD at a temperature of $2500$K for $10$ps. Five random states were selected along each of the 37 trajectories (blue dots in figure~\ref{fig:scatterPlot}). Then each state was relaxed with DFT-GGA computations at fixed volumes producing strings of intermediate energy states (red dots in figure~\ref{fig:scatterPlot}). For clarity only two strings are shown in the figure, at $\beta=0.5, 0.71$, the rest of the red dots show the final states of each string. The yellow region shows the configuration space that the intermediate states cover. The database is comprised of 44,596 states, and we note that the precision of the Li-Si training set is based on a DFT precision of 1 meV/atom. For a complete account of this database see \cite{kim:thesis2018}.  

\section{Numerics}
\label{sec:numerics}

\subsection{Model and Training Method} \label{sec: model}

Recall that each state is specified by $x=\{ (z_k,r_k) \}_{k=1}^N$ for $z_k$ the total protonic charge of the atom at $r_k\in\mathbb{R}^3$. We define a naive electronic density $\rho_x$ as $\rho_x(u) = \sum_{k} z_k g(u-r_k)$ with centered and normalized Gaussians $g(u)=(\sqrt{2\pi}\alpha)^{-3} e^{-\abs{u}/2\alpha^2}$. Therefore $\| \rho_x \|_1 = \sum_k z_k$ is the total protonic charge in the unit cell (periodic box) of the state $x$. Rather than use $\rho_x$ as a single channel input to the scattering transform, we instead partition the state information into five density channels corresponding to the core electrons of the system, the valence electrons of the system, the Li atoms only, the Si atoms only, and the square root of the full system (to capture the kinetic energy). We denote these channels by $\rho_x^{\delta} (u)$, where $\delta \in \{ \text{core}, \text{valence}, \text{Li}, \text{Si}, \text{atomic sqrt} \}$.

In this work our focus is the predictive capabilities of the wavelet scattering coefficients, and so we perform a simple linear regression over the scattering coefficients as
\begin{equation}
    \widetilde{E}_f(x) = w_0 + \sum_{\delta, \lambda_1, q} w_{\delta \lambda_1 q}\frac{\norm{\sigma(\rho_x^{\delta} \circledast \psi_{\lambda_1})}_q^q}{\norm{\rho_x^{\delta}}_q^q} + \sum_{\delta, \lambda_1, \lambda_2, q} w_{\delta \lambda_1 \lambda_2 q} \frac{\norm{\sigma(\sigma(\rho_x^{\delta} \circledast \psi_{\lambda_1}) \circledast \psi_{\lambda_2})}_q^q}{\norm{\sigma(\rho_x^{\delta} \circledast \psi_{\lambda_1})}_q^q}, \label{eqn:regression}
\end{equation}
subject to $\| w \|_0 \leq M$, with wavelet parameters $\lambda=(n,l,j)$ (using the atomic orbital wavelets of example \ref{ex:atomic orbital wavelets}) and where we select $M$ non-zero weights $w = \{w_0, w_{\delta \lambda_1 q}, w_{\delta \lambda_1 \lambda_2 q}\}$ using greedy orthogonal least squares regression \cite[Sec. 3.2]{hirn:waveletScatQuantum2016}. For small $M$ we thus obtain a sparse regression, which can lead to an interpretable model.

\subsection{Numerical Results and Analysis}

While there is potential for using the scattering coefficients to predict many properties, the aim for this paper is to predict the formation energy to high accuracy. Motivated by the precision of DFT, we seek to obtain a regression accuracy of 1.0 meV/atom relative to the DFT computed energies.

To robustly test our model against all points in the data set, we used five-fold cross validation. Folds were selected randomly. In numerical computations we used the five density channels described in section \ref{sec: model}, six wavelet scales increasing by factors of $\sqrt{2}$, atomic orbital wavelets for $1 \leq n \leq 5$ and all valid choices of $l$, and $q = 1, \nicefrac{4}{3}, \nicefrac{5}{3}, 2$. Second order scattering coefficients were restricted so that $j_2 > j_1$ and $(n_2, l_2) > (n_1, l_1)$ (by dictionary ordering). The hyperparameter $M$, indicating the number of scattering coefficients to use in the linear model \eqref{eqn:regression}, was selected using cross validation on the training set. Bagging was used to learn multiple models for each test set, and the final predicted formation energy was taken as the average of the predicted energies of each model.

\begin{table}[hbtp]
\caption{Formation energy regression errors on Li-Si test data measured in meV/atom. Note that while training and test data sets are qualitatively similar across the methods, they are not the same and thus the average errors presented here are not directly comparable.}
\centering
\begin{tabular}{| l l l l l |}
    \hline
     & Artrith \cite{Artrith:LiSiML2018} & Onat \cite{onat:implantNNLiSi2018} & First order scattering & First \& second order scattering \\
    \hline
    MAE & 5.9   & -- & 2.13 $\pm$ 0.02   & 0.78 $\pm$ 0.01   \\
    \hline
    RMSE & 7.7   & 4.5   & 2.95 $\pm$ 0.04   & 1.24 $\pm$ 0.05  \\
    \hline
\end{tabular}
\label{tab: regression comparison}
\end{table}

\begin{wrapfigure}[21]{R}{0.3\linewidth}
  \centering
  \includegraphics[trim={0in 0.15in 0in 0.4in},clip,width=0.99\linewidth,]{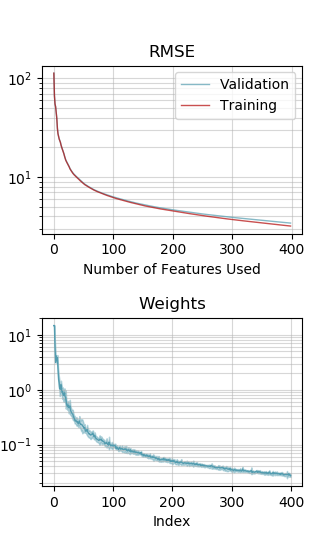}
  \caption{Top: Regression RMSE vs. model size. Bottom: weights vs. number of training points.}
  \label{fig:rmse and training}
\end{wrapfigure}

Utilizing only first order scattering coefficients in the linear model \eqref{eqn:regression}, we obtain a mean absolute error (MAE) of 2.1 meV/atom on the testing folds, which surpasses the results of neural network models trained and tested on similar Li-Si databases as reported in \cite{Artrith:LiSiML2018, onat:implantNNLiSi2018}. We note, however, that the data in \cite{Artrith:LiSiML2018, onat:implantNNLiSi2018} is not currently available, and thus this is not a precise comparison. In order to reduce the regression error, we also trained the linear model \eqref{eqn:regression} using both first and second order scattering coefficients. With these additional coefficients, the linear regression model \eqref{eqn:regression} achieved an MAE of 0.78 meV/atom with a standard deviation across the five folds of 0.01 meV/atom. Table \ref{tab: regression comparison} summarizes these results.

\begin{wrapfigure}[15]{R}{0.45\linewidth}
    \centering
    \includegraphics[trim={0.7in 0in 0.7in 0.68in},clip,width=0.98\linewidth]{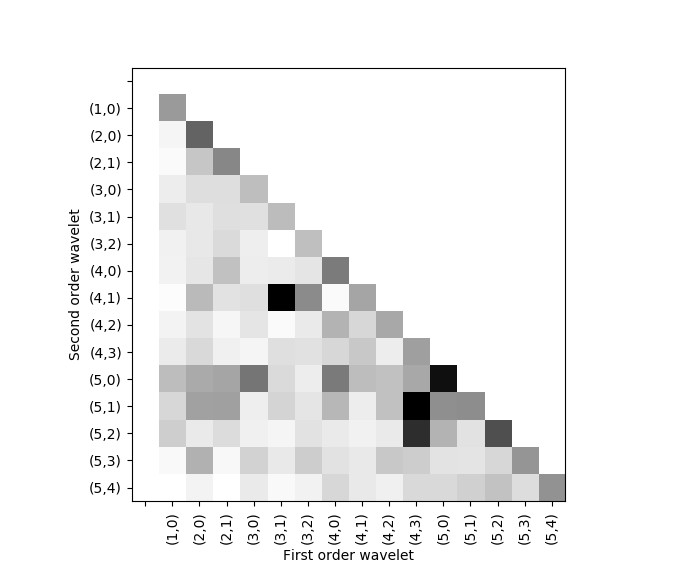}
    \caption{Sum of model weights. Darker squares indicate larger weights.}
    \label{fig:weight sums}
\end{wrapfigure} 

Our results display several expected machine learning phenomena that indicate the scattering coefficients are capturing the structural variability in the Li-Si database. 
The top of figure \ref{fig:rmse and training} plots the RMSE as a function of $M$, the number of scattering coefficients used in the linear model \eqref{eqn:regression}. It shows the error decays steeply at first before flattening, which indicates the first few selected features are highly correlated with the formation energy; one can thus obtain a very sparse (i.e., simple) model that still obtains good accuracy (e.g., $M = 100$ scattering coefficients obtains an RMSE of $5.8$ meV/atom). Also of importance, the validation curves maintain close proximity to the training curves, which suggests that we are learning more than the noise of the training set. Of particular note in figure \ref{fig:rmse and training} is the slope of the error curve remains negative, meaning there is a potential to extend the model using additional scattering coefficients and improve the accuracy further.

\begin{figure}[t]
    \centering
    \includegraphics[trim={1.2in 0.0in 1.2in 0.45in},clip,width=1.\textwidth]{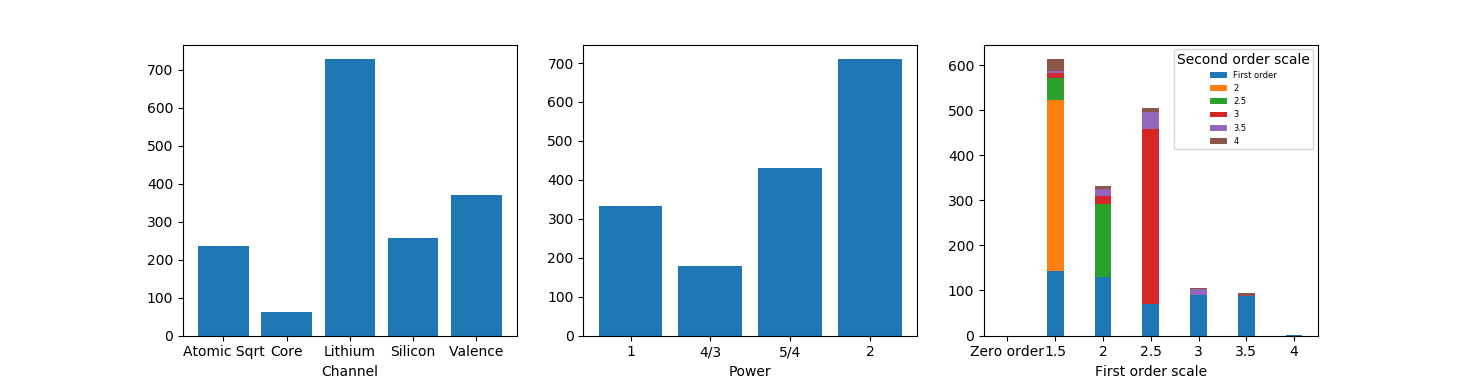}
    \label{fig:stats bars}
    \vspace{-10pt}
    \caption{Statistics of selected wavelet coefficients. From left to right: histogram by channel $\delta$; histogram by power $q$; histogram by scales $(j_1, j_2)$.}
\end{figure}

The greedy selection process enables us to extract an ordering of the scattering coefficients which were most critical in predicting the energy. Figure \ref{fig:rmse and training} (bottom) plots the corresponding weights of the scattering coefficients according to their selection index; we see a rapid decay, again indicating a sparse model. We also compute statistics of the selected features to get a broader picture of the process. The two plots on the left of figure \ref{fig:stats bars} plot the usage of scattering coefficients in the model \eqref{eqn:regression} by channel $\delta$ and power $q$, respectively, according to weights $w$. From these we see that coefficients derived from the Li-only channel were used most by the regression, and similarly for the power $q = 2$. Interestingly, the power $q = \nicefrac{4}{3}$ was used the least; these terms correspond to the exchange energy, which constitutes a small fraction of the total energy. Additionally, the valence channel is favored over the core channel, matching the larger role of valence electrons in bond formation.

The rightmost plot of figure \ref{fig:stats bars} shows similar information for scales. The empty first space represents the bias term $w_0$. The blue portions represent first order features and all other colors are second order scattering coefficients. As indicated by the orange, green, and red bands in the second, third, and fourth columns, the model prefers to couple wavelets of similar scales. Figure \ref{fig:weight sums} plots the sum of the weights corresponding to the wavelets, where the diagonal represents first order wavelet coefficients. We can get an idea of which wavelets are used most frequently, but no clear pattern emerges. 

\section{Conclusion} \label{sec: conclusion}

We have presented a general family of wavelet scattering coefficients as features for use in machine learning tasks on three dimensional signals. The features are specially suited for modeling properties of complex multiscale materials; they have desirable invariant properties, including invariance to rigid translations, rotations, and reflections. Additional invariants, such as to system size, are recovered through normalization of the scattering coefficients. These features may be tailored for the application at hand through the selection of the radial component of the wavelets \eqref{eq:generalwavelet}. General considerations for selecting the radial components and an example selection were presented in example~\ref{ex:atomic orbital wavelets}. These wavelets give state-of-the-art results for modeling formation energies of amorphous Li-Si systems. The numerics show that an interpretable (sparse) model is recovered which appears to be improvable with the inclusion of additional features. 
The computed model performed admirably when compared with two other methods trained on similar Li-Si databases. Future work will include investigations into alternative choices of radial functions for the proposed wavelets as well as the effectiveness of these features in modeling other material properties and signals.

\subsubsection*{Acknowledgments}

We would like to acknowledge Mr. Jialin Liu's contribution to automate the training set generation. X.B., P.S., K.J.K. and M.H. were (partially) supported by DARPA YFA \#D16AP00117. M.H. is also supported by Alfred P. Sloan Fellowship \#FG-2016-6607 and NSF grant \#1620216. Y.Q. also appreciates the financial support from Michigan State Research Foundation (SPG grant).

\small

\bibliographystyle{unsrt}
\bibliography{MainBib}

\end{document}